\documentstyle[11pt,epsfig]{article}
\begin{document}

\begin{center}

{\bf \Large
On the chaoticity of active-sterile neutrino oscillations
in the early universe}
\bigskip
\\{\bf Poul-Erik N. Braad
\footnote{e-mail: {\tt peb@ifa.au.dk}}\\
{\small {\it{Institute of Physics and Astronomy,
University of Aarhus, 
DK-8000 \AA rhus C, Denmark
\\
}}}}
{\bf Steen Hannestad \footnote{e-mail: {\tt steen@nordita.dk}} \\
{\small {\it{Nordita, Blegdamsvej 17, DK-2100 
Copenhagen, Denmark
\\
}}}}
\end{center}

\begin{abstract}
We have investigated the evolution of the neutrino asymmetry in
active-sterile neutrino oscillations in the early universe.
We find that there are large regions of parameter space where
the asymmetry is extremely sensitive to variations in the initial
asymmetry as well as the external parameters (the mass difference
and the mixing angle). In these regions the system undergoes
chaotic transitions; however, the system is never truly chaotic
in the sense that all information about initial conditions is lost.
In some cases though, enough information is lost that the final sign
of the neutrino asymmetry is stochastic.
We discuss the implications of our findings for Big Bang 
nucleosynthesis (BBN) and the cosmic microwave background (CMB).
\end{abstract}


Neutrino oscillations have been proven beyond reasonable doubt by
the observation of the solar and atmospheric neutrino anomalies.
The solar neutrino problem can be explained by oscillations between
$\nu_e$ and $\nu_\mu$ \cite{Bahcall}, whereas the atmospheric neutrino deficit 
is quite nicely explained by oscillations between $\nu_\mu$ and
$\nu_\tau$ \cite{Fukada}. However, the LSND experiment also claims detection
of neutrino oscillations between $\nu_\mu$ and $\nu_e$, but with
a much larger mass difference than found from the solar neutrino
experiments \cite{lsnd}. There is no possible three-neutrino solution to the
combined data. Either one of the interpretations is wrong, or there 
is a fourth neutrino species, responsible for either the atmospheric
or the solar neutrino anomaly.
It has proven difficult to rule out the existence of such a fourth
neutrino. It cannot interact via the usual weak interactions, or it
would have been detected in $Z$-decay experiments.
However, such additional light sterile neutrinos are predicted to exist in
many extensions of the standard model, and it is certainly worthwhile
to study the implications of such neutrinos for cosmology.

Active-sterile neutrino oscillations in the early universe have been
intensely studied for more than a decade. The pioneering studies concentrated on effects from
excitation 
 of inert degrees of freedom from the background plasma on
BBN predictions of the abundances of the light elements 
[4-10]. Extra degrees of freedom exited before
the BBN epoch increases the relativistic energy density of the
universe. This modifies the expansion rate of the universe and changes
the outcome of BBN, most notably the primordial helium abundance, $Y_P$.

It was realized early on that neutrino oscillations in a CP asymmetric
plasma like the early universe could amplify an initial
neutrino asymmetry $(L_\nu \equiv (N_\nu-N_{\bar{\nu}})/N_\gamma)$
\cite{Barbieri}. Initially, the effect was thought unimportant
\cite{EKM1}, and it was not until 1996 where Foot, Thomson and Volkas
\cite{FTV96} showed that for for some
mixing parameters, an important and non negligible
asymmetry can actually be generated and affect BBN-predictions \cite{FTV96,lbbn}. These results were confirmed by
Shi \cite{shi}, who in addition argued that in parts of the $(\delta m^2,\sin 2\theta)$-parameter space, the sign of the
asymmetry can go though a period of rapid oscillations
before settling down at a final value. The magnitude of the final
asymmetry seems quite robust, but the final sign is very sensitive to
small changes in the initial asymmetry and the mixing
parameters. These results have later been investigated in further
detail.

Based on the quantum rate equations (QRE's), Enqvist et
al. \cite{enqvist} have confirmed
Shi's results
The set of QRE's is an approximation
to the full quantum kinetic equations (QKE's) for the system,
based on the assumption that all modes have the average momentum,
$p = \langle p \rangle \simeq 3.15 T$.
This approximation is excellent outside resonance regions, but its
applicability inside resonances has been questioned. The reason is
that in the QKE's different momentum modes pass through the resonance
at different temperatures. Assuming only one mode could possibly
lead to different behaviour of the oscillations inside the resonance.

Unfortunately it is not yet possible to solve the full QKE's through
a strong resonance due to numerical limitations \cite{barifoot,wong}. This fact has 
lead to recent controversy\footnote{The controversy raised
by Dolgov et al. \cite{hansen} seems to have been resolved and the
conclusions of Dolgov et al. rejected (see Refs. \cite{sorri}).} about the behaviour of the asymmetry 
when the full set of QKE's is employed.
We have chosen to work with the quantum rate equations in the present
treatment because the equations are numerically solvable and reliable
conclusions can be drawn from them. However, we caution that some of
our conclusions do not necessarily apply to the full set of quantum
kinetic equations.

The question of chaoticity is important since chaotic generation of
lepton number might lead to the formation of leptonic domains. Not
causally connected regions could develop different signs on the
leptonic asymmetry. Diffusion of neutrinos over leptonic boundaries
would be expected to suppress asymmetry and might also open a new
channel of producing sterile neutrinos via resonant MSW conversion of
active neutrinos at domain boundaries \cite{domain}

Our goal is to investigate to what extend the final sign of the
asymmetry can be stochastic. In order to do this we employ a
Lyapunov-exponent analysis of the system and calculate the
loss of information
about initial conditions experienced as the system passes through the
resonance.
Finally, we discuss the implications of our findings for BBN
and the CMBR.

{\it Equations ---}
The two-neutrino mixing is characterised by the vacuum
mixing angle $\theta$. The matter eigenstates $\nu_{1,2}$ are connected to the
flavor eigenstates $\nu_{\alpha,s}$ via the
transformations\footnote{Note, the absolute phase of the non-diagonal
  terms is arbitrary; our choice differs from some in the literature.}
\begin{eqnarray}
\label{a.0.0.1}
\nu_{\alpha} & = & \cos \theta \, \nu_{1}-\sin \theta \,
\nu_{2}\nonumber \\
\label{a.0.0.2}
\nu_{s} & = & \sin \theta \, \nu_{1}+\cos \theta \, \nu_{2}.
\end{eqnarray}
If the two mass eigenstates have masses $m_{1}$ and $m_{2}$, the
squared mass difference is defined as; $\delta m^{2}=m_{2}^{2}-m_{1}^{2}$.
The objects of interest are the reduced 2$\times$2-density matrices
$\rho_{\nu}$ and $\rho_{\overline{\nu}}$ for the neutrino and
anti-neutrino ensembles respectively. Expanded in \cite{EKM1,McKThom}
Pauli spin matrices these take the form 
\begin{equation}
\label{a.0.0.3}
\rho_{\nu}=\frac{1}{2}P_{0}(1+\mathbf{P}\cdot \sigma); \qquad 
\rho_{\overline{\nu}}=\frac{1}{2}\overline{P}_{0}(1+\mathbf{\overline{P}}\cdot
\sigma).
\end{equation}
The coefficients in the expansion form the components of the
polarisation vector $\mathbf{P}$. 

The diagonal elements in  $\rho_{\nu}$ are the relative number
densities of the active neutrino and the sterile neutrinos normalised
to their equilibrium values 
\begin{equation}
  \label{a.0.0.4}
  n_{\nu_{\alpha}}=\frac{1}{2}P_{0}(1+P_{z}), \qquad n_{\nu_{s}}=\frac{1}{2}P_{0}(1-P_{z})
\end{equation}
with similar equations for the antineutrino ensemble. With the adopted
normalisation, $P_{0}$ then equals the sum of the relative number
densities of the mixed neutrinos
(2 in chemical equilibrium).
$P_{z}$ measures the distribution between the
active and the sterile neutrinos. If $P_{z}$=1 all neutrinos in the
mixed ensemble will be of the active type, whereas all neutrinos will
be of the sterile type if $P_{z}=-1$.

Instead of solving the full set of QKE's for $\rho_\nu(p)$ and $\rho_{\bar{\nu}}(p)$,
we solve the quantum rate equations for $\rho_\nu(p = \langle p \rangle)$ and 
$\rho_{\bar{\nu}}(p = \langle p \rangle)$, where $\langle p \rangle \simeq 3.15 T$ 
\footnote{One can also argue that the momentum should be $p \simeq 2.1 T$,
which is the maximum for the Fermi-Dirac distribution function.
Taking this value instead does not make any qualitative difference, but
can make some quantitative difference because the resonance temperature
is altered by a factor $(3.15/2.1)^{1/2} \simeq 1.22$.}.

We focus on the part of parameter space where the the final sign of the
leptonic asymmetry appears extremely sensitive towards small changes
in the parameters. In this part of parameter space collision terms can
be neglected so that the resonance is non-adiabatic and no significant
amount of sterile neutrinos are produced. $P_{0}$ is then a constant
which can be put equal to unity. The approximation is valid for
mixing parameters obeying the inequality \cite{ShiFie}
\begin{equation}
\label{a.0.0.7}
|\delta m^{2}|\sin ^{4}2\theta <10^{-9}.
\end{equation}
The coupled equations of motion then are
\begin{equation}
\label{a.0.0.8}
\dot{\mathbf{P}}=\mathbf{V}\times \mathbf{P}-D\cdot \mathbf{P_{T}},
\qquad \dot{\mathbf{\overline{P}}}=\mathbf{\overline{V}}\times
\mathbf{\overline{P}}-\overline{D}\cdot \mathbf{\overline{P}_{T}},
\end{equation}
where $\dot{\mathbf{P}}\equiv d\mathbf{P}/dt$, and the transversal
polarisation vector $\mathbf{P_{T}} \equiv$
$P_{x}$ $\hat{\mathbf{x}}+P_{y}\hat{\mathbf{y}}$ has been introduced. For
definiteness we shall here focus on $\nu_{\tau}-\nu_{s}$ oscillations;
other cases are obtained from this by simple redefinitions. The
damping coefficient $D$ signifies quantum damping introduced in the
system by coherent and incoherent scattering of
neutrinos on particles in the background plasma. In the case of
$\nu_{\tau}-\nu_{s}$ oscillations it is given by  \cite{EnKaThom,McKThom}
\begin{equation}
\label{a.0.0.9}
D\simeq \overline{D} \simeq 1.83\,G_{F}^{2}T^{5}\simeq 2.49 \times 10^{-46}\,T^{5}\textrm{ eV}^{-4}.
\end{equation}

The rotation or potential vector $\mathbf{V}$ arises from an expansion
of the total weak Hamiltonian in Pauli spin matrices. In the case of
$\nu_{\tau}-\nu_{s}$ oscillations it is given by
\begin{equation}
\label{a.0.0.10}
\mathbf{V}=-b_{s} \,\mathbf{\hat{x}}+(-b_{c}
-V_{\nu_{\tau}})\,\mathbf{\hat{z}},\qquad \mathbf{\overline{V}}=-b_{s}
\,\mathbf{\hat{x}}+(-b_{c}
-\overline{V}_{\nu_{\tau}})\,\mathbf{\hat{z}},
\end{equation}
where
\begin{equation}
\label{a.0.0.11}
b_{s}=-\frac{\sin 2\theta \, \delta m^{2}}{2\langle p \rangle} \simeq
  -\frac{\sin 2 \theta \, \delta m^{2}}{6.3 \, T},    
\end{equation}
and
\begin{equation}
\label{a.0.0.12}
b_{c}=-\frac{\cos 2\theta \, \delta m^{2}}{2\langle p \rangle} \simeq
  -\frac{\cos 2 \theta \, \delta m^{2}}{6.3 \, T}. 
\end{equation}
The term $V_{\nu_{\tau}}$($\overline{V}_{\nu_{\tau}}$) is an effective
energy contribution to the $\nu_{\tau}$($\overline{\nu}_{\tau}$) from
neutral current and charged current interactions with particles in the
background. It is given by \cite{raff,EnKaThom}
\begin{equation}
\label{a.0.0.13}
V_{\tau }=-b_{T}+b_{A}\delta P_{z},\qquad \overline{V}_{\tau
  }=-b_{T}-b_{A}\delta P_{z}
\end{equation} 
The parameter $b_{T}$ is the contribution to the effective potential
from the low energy tail of the vector bosons
\begin{equation}
\label{a.0.0.14}
b_{T}=\frac{4\sqrt{2}\xi (3)}{\pi ^{2}}\frac{\cos ^{2} \theta _{W}
  \langle p \rangle ^{2}}{M_{W}^{2}}G_{F}T^{3}=9.36\times 10^{-45}\
\textrm{eV}^{-4}\,T^{5}.
\end{equation}
The second term in Eq. (\ref{a.0.0.13}) is a matter potential arising
in an asymmetric background. The parameter $b_{A}$ is given by
\begin{equation}
\label{a.0.0.15}
b_{A}=\frac{3\sqrt{2}\xi (3)}{4\pi ^{4}}G_{F}T^{3}\simeq 1.53 \times
10^{-24} \ \textrm{eV}^{-2}\,T^{3}
\end{equation}
and we have defined the parameter 
\begin{equation}
\label{a.0.0.16}
\delta P_{z}=\frac{8}{3}L^{(\tau)}.
\end{equation}
The term $L^{(\mu ,\tau )}$ contains all the asymmetries in the problem
\begin{equation}
\label{a.0.0.17}
L^{(\tau)}=2L_{\nu_{\tau}}+L_{\nu_{\mu}}+L_{\nu_{e}}-\frac{1}{2}L_{n}.
\end{equation}
An asymmetry is defined as
$L_{\nu_{\alpha}}=(N_{\nu_{\alpha}}-N_{\overline{\nu}_{\alpha}})/N_{\gamma}$,
where $N_{\nu_{\alpha}}=(3/8)N_{\gamma}n_{\nu_{\alpha}}$ is the number
density of a given neutrino specie and $N_{\gamma}=2\xi (3)T^{3}/\pi
^{2}$ is the photon number density. Specifically for $\nu_{\tau}$ one
finds
\begin{equation}
\label{a.0.0.19}
L_{\nu_{\tau}}=\frac{3}{8}(P_{z}-\overline{P}_{z}),
\end{equation}
such that 
\begin{equation}
\label{a.0.0.20}
\delta P_{z}=P_{z}-\overline{P}_{z}+P_{c},
\end{equation}
where $P_{c}=(8/3)L^{\rm in}$ is a constant initial asymmetry.

The expansion of the universe is given by the Friedmann equation
$H = \sqrt{8\pi G \rho/3}$, where $\rho = \frac{\pi^2}{30} g_* T^4$.
In accordance with our assumption of non-adiabatic oscillations
we take $g_* = Const = 10.75$ \cite{kolb}.

Since we are focusing on the evolution of small asymmetries it is an
advantage to redefine the polarisation vectors ($\mathbf{P}$ and
$\mathbf{\overline{P}}$) as
\begin{equation}
\label{a.0.0.24}
P_{i}^{\pm}=P_{i}\pm \overline{P}_{i}\qquad (i=x,y,z).
\end{equation}
in order to separate small $(P^-)$ and large $(P^+)$ quantities.
This greatly increases the numerical accuracy.

{Numerical results ---}
We have numerically studied the system of equations at temperatures
below 100 MeV. At high
temperatures the $\nu_\tau$ will be in thermal equilibrium via its
weak interactions. The $\nu_s$ can only be brought
into thermal equilibrium by oscillations. However, at high
temperatures the damping is very strong and oscillations therefore
totally washed out. Initially the abundance of $\nu_s$ will therefore
be practically zero. 
The initial lepton asymmetry is plausible comparable in magnitude
to the baryon asymmetry $(L^{\rm in} \sim 10^{-10}-10^{-9}$ (however,
this need not be the case).
The evolution of the leptonic
asymmetry depends upon the sign of $\delta m^2$. If $\delta m^2 <0$
the system encounters a MSW resonance where the effective matter
mixing angle becomes maximal at a 
temperature (see e.g. \cite{shi})
\begin{equation}
\label{1}
T_{\rm res} =16.0\left( \cos 2\theta \cdot (|\delta m^2 |/\textrm{eV}^{2}) \right) ^{1/6} \textrm{ MeV}.
\end{equation}
We find solutions very similar to those discussed by Enqvist et
al.\cite{enqvist}. We also find that in large regions of parameter
space the sign of the final asymmetry
is extremely sensitive to small variations in mixing parameters and
initial asymmetry. It is interesting to investigate if the observed
sign indeterminacy is a real chaotic phenomenon in the mathematical
sense that all information about the initial condition is lost during
the process of asymmetry amplification or if the system is just
``sensitive''.

In order to investigate these dynamical aspects of the system,
we have numerically calculated the Lyapunov characteristic exponents, $\lambda_i$
(LCE's). The LCE's measure the exponential divergence and convergence
of nearby orbits in phase space. A dissipative system with a positive
LCE is defined as chaotic since it ultimately losses all information
about the initial condition. A dissipative system contracts phase
space volume in time $(\sum \lambda_i < 0)$.
The set of QRE's is a dissipative dynamical system because of the
expansion of the universe, so that the LCE's is an ideal way to 
study the properties of the system.

$\lambda_i >0$ is not only a necessary condition for chaos, it also
quantifies the degree of chaos in the system.
Its magnitude is a measure of how long
it takes for an initial uncertainty do develop into total
chaos. The information loss $I$ related to a postive LCE is basicly after a time $t$  equal to $I=\lambda t$ and is typically measured in bits.  
  
We have numerically calculated the LCE's for the QRE's (\ref{a.0.0.8}). The
calculation has been based on a method by Benettin \cite{wolf,Benettin}. 
An orbit in
6-dimensional phase-space is defined by integrating the non-linear
system of 6 QRE's. Surrounding an initial point on this reference
orbit is fixed a sphere defined by 6 principal axes. The surface of
this sphere now defines the different initial conditions in the
problem. When integrating the non-linear system forward in time the
sphere follows the point on the reference orbit that it is fixed
to. While doing this it becomes deformed to a 6-ellipsoid because of
the divergence or convergence of close orbits in phase-space. We define
the LCE's from the change in the magnitude of the lengths $\vert
\vec{v}_{i} \vert$ of the spheres' principal axes

\begin{equation}
\label{6.2.2.6}
\lambda_{i}=\lim_{t \to \infty}\, \frac{1}{t}\, \log_{2}\frac{\vert \vec{v}_{i}(t) \vert }{\vert \vec{v}_{i}(0) \vert },
\end{equation}
The evolution of the sphere's
principal axes is defined by a linearised version of the non liniar
set of equations. In each integration point the magnitude of the
sphere's principal axes are compared to the magnitude at the
previous integration point. Averaged
over time this produces the time averaged LCE's \cite{wolf}.

The system of 6 QRE's contains 6 different LCE's. Until the system hits
the resonance all 6 exponents are negative and no information
lost. Once the system encounters the resonance, the leptonic asymmetry
begins to oscillate rapidly from positive to negative values. At this
point the dynamics of the system changes; it now has 1 positive and 5
negative exponents, so information is lost. Fig. 1 has been produced
form the initial conditions ($\delta m^2$, $\sin^2 2\theta$,
$P_c$)=($-10^{-3} \textrm{ eV}^2$, $10^{-5}$, $10^{-10}$). The upper
panel shows the asymmetry as a function of temperature
inside the resonance, whereas the lower displays the loss of
information through the resonance. This information loss has been
averaged over 100 different initial conditions which lie very close
in phase space. This eliminates noise and does not in any way distort the
conclusions.

The LCE's in general depend upon the initial conditions. Only if an
ergodic measure exists for the attractor, will almost all initial
conditions in the basin of attraction result in the same LCE's
\cite{Ruelle}. Often, 
the structure of the basin of attraction is so
complicated that even a small variation in initial conditions puts one
in the basin of attraction of another attractor, characterized by a different
set of LCE's~\cite{schuster}.
\begin{figure}[t]
\begin{center}
\epsfig{file=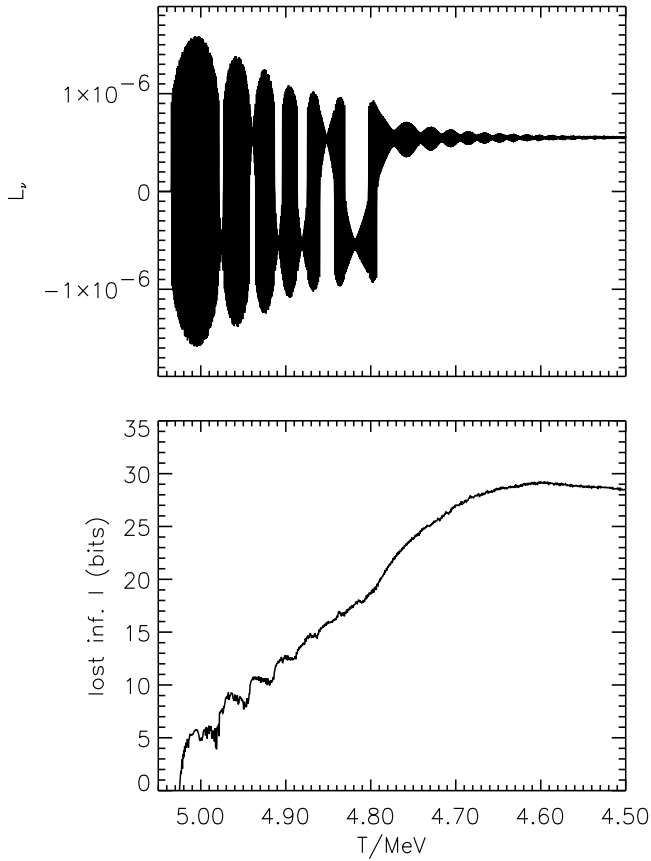,height=11cm,width=8.5cm}
\end{center}
\vspace*{0.5cm}\hspace*{1cm}
\caption{Upper panel: The evolution of $L_\nu$ with temperature for the parameters
$\delta m^2 = -10^{-3} \, {\rm eV}^2$, $\sin^2 2 \theta = 10^{-5}$ and $L_\nu^{\rm in} = 10^{-10}$.
Lower panel: The total loss of information in bits for the same model.}
\label{fig2}
\end{figure}

In Fig. 1 it seems reasonable that the situation with a
negative final asymmetry can not be characterized by the same attractor as
the situation with a positive final asymmetry.
Most likely a different type of attractor also appears when the system
crosses the resonance, namely a {\it chaotic} attractor \cite{ott}.
Since
different attractors in general are characterized by different LCE's,
one must average the LCE's over different initial conditions to probe
the spatial structure of the attractor.

From Fig. 1 it is noticed that as the asymmetry initially grows and
starts oscillating rapidly, information is lost. After the initial violent
growth the lepton number oscillates more regularly, and information is no longer lost. Very suddenly
the system begins then to oscillate with much smaller amplitude around one of the two
attractors. This can be considered a chaotic transition where, as
displayed in the lower part of Fig. 1, information is rapidly
lost. The system goes through this alternating behaviour of "ordinary
oscillations`` followed by chaotic transitions all the way trough the
resonance\footnote{For larger $\delta m^2$ the system only goes
  through chaotic transitions at the beginning of the
  resonance. Afterwards the asymmetry oscillates with slowly
  increasing asymmetry until the end of the resonance. This is the
  behaviour seen in Fig. 1 of Ref. \cite{enqvist}. However,
  information is only lost when the chaotic transitions are
  present.}. 
However, ultimately the system is driven out of the
chaotic region by the expansion of the universe such that only a limited amount
of information is lost. 
\begin{figure}[t]
\begin{center}
\epsfig{file=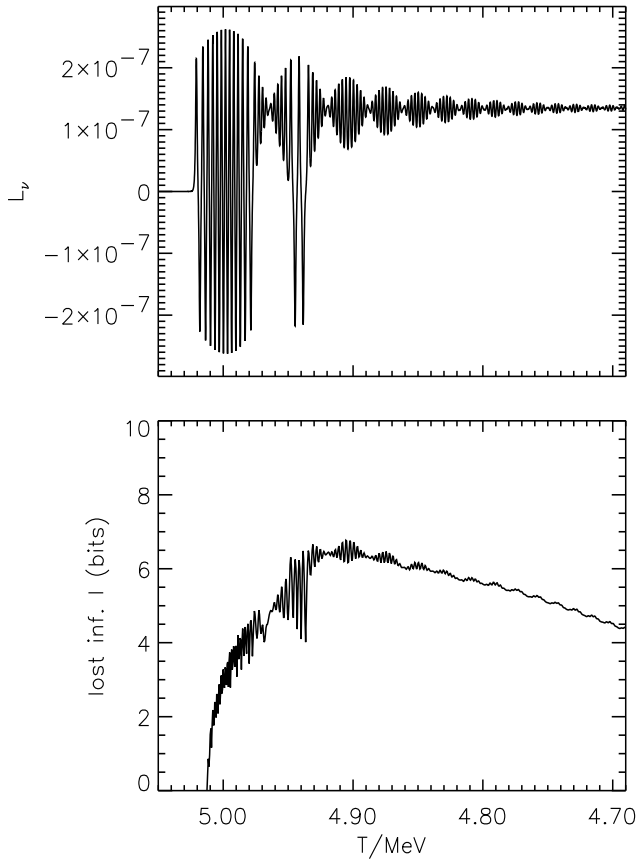,height=11cm,width=8.5cm}
\end{center}
\vspace*{0.5cm}\hspace*{1cm}
\caption{Upper panel: The evolution of $L_\nu$ with temperature for the parameters
$\delta m^2 = -10^{-3} \, {\rm eV}^2$, $\sin^2 2 \theta = 10^{-7}$ and $L_\nu^{\rm in} = 10^{-10}$.
Lower panel: The total loss of information in bits for the same model.}
\label{fig1}
\end{figure}

The system is not truly chaotic in the sense that it has a positive
time averaged Lyapunov exponent at all times. However, it goes trough
the chaotic region and will therefore be sensitive to
variations in initial conditions. This sensitivity depends on the
amount of information lost, which again depends
critically on the value of the mixing parameters. 
Fig. 2 has been produced from the initial
conditions ($\delta m^2$, $\sin^2 2\theta$,
$P_c$)=($-10^{-3} \textrm{ eV}^2$, $10^{-7}$, $10^{-10}$) and
illustrates nicely that the loss of information is smaller for smaller
mixing angles.

In general we find
that the
sensitivity grows with larger mixing angles and smaller mass
differences\footnote{This conclusion is supported by Fig.2 of Ref.~\cite{enqvist}.}. 
This is clearly seen in Fig.~3 where we have calculated the total loss of information
(in bits)
for different values of $\delta m^2$ and $\sin^2 2 \theta$.

\begin{figure}[htb]
\begin{center}
\epsfig{file=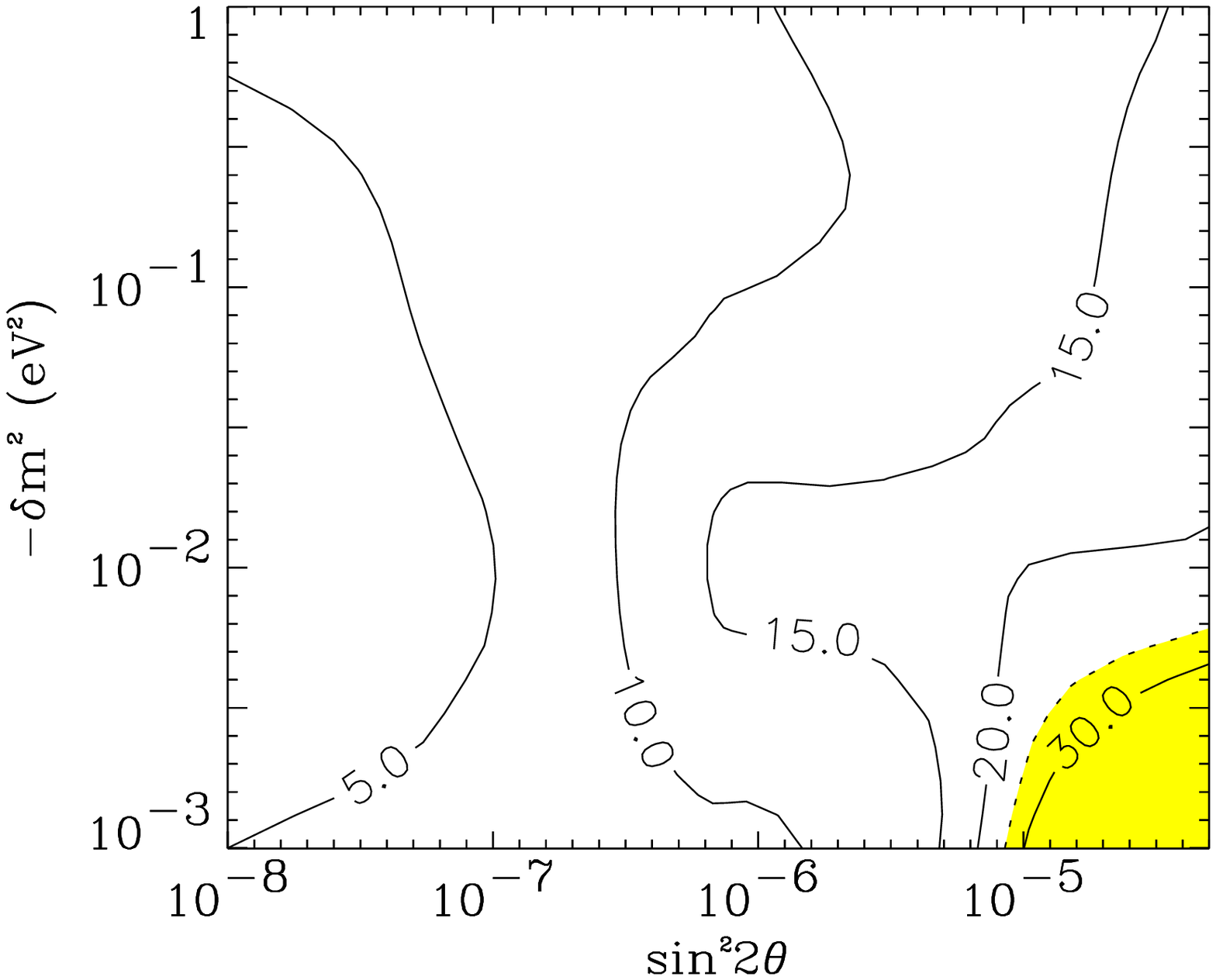,height=8cm,width=10cm}
\end{center}
\vspace*{0.5cm}\hspace*{1cm}
\caption{The loss of information, calculated for different values of $\delta m^2$ and
$\sin^2 2 \theta$. In all cases we have assumed an initial asymmetry of $L_\nu^{\rm in} = 10^{-10}$.}
\label{fig3}
\end{figure}

{\it Discussion ---}
We have performed a careful study of the dynamical properties of the
quantum rate equations for active-sterile neutrino oscillations in
the early universe. Numerically we have confirmed the results of
Ref.~\cite{enqvist}, that there are regions of parameter
space where the final sign of the lepton asymmetry is extremely sensitive
to initial conditions. However, even though the systems goes through
chaotic transitions, it is never truly chaotic in a mathematical sense
because the system only stays a finite time in the chaotic regions.

Effectively though, the system can behave exactly like a chaotic system in
some regions of parameter space. Thermal fluctuations in the early 
universe can be expected to produce slightly different initial
asymmetries, of the order $\Delta L^{\rm in} \simeq 10^{-18}$
\cite{barifoot}. If the 
average initial asymmetry is of the order $L^{\rm in} \simeq 10^{-10}$
the information loss needs only be of the order 25 bits in order
to produce a completely arbitrary final sign. We have indicated this region of 
$(\delta m^2,\sin ^2 2 \theta)$-space in Fig.~3 by a light shading.

It should be noted that there are reasons for believing that much
larger initial differences in $L$ could be present. They could for instance
have been produced during the electroweak phase transition (there could
be cases where $\Delta L^{\rm in} \simeq L^{\rm in}$). In that case, large
regions in the mixing parameter space leads to arbitrary sign, essentially
the asymmetry only has to go through a few oscillations.

The outcome of BBN depends strongly on the sign of a possible lepton
asymmetry, if the asymmetry is in the electron neutrino sector. 
The reason is that electron neutrinos participate directly in the weak
reactions that interconvert neutrons and protons. A positive $L$ leads
to an decrease in $Y_P$ and a negative $L$ to an increase. So in principle
different regions could have different light element abundances. The
uncertainty in $Y_P$ is expected to be of the order $10^{-2}$.
The formation of domains with different lepton number can in principle
also lead to visible effects in the CMB. 
The extra energy density produced by repopulation is in most cases
very small (which is exactly why our non-adiabatic approximation applies),
so the oscillations are not detectable in this way \cite{cmbr}. However,
the recombination history depends sensitively on $Y_P$, so the CMB
anisotropies are still altered to some extent by the oscillations.

Finally, we caution that our calculations have been performed using the
quantum rate equations (QRE's) instead of the full set of quantum kinetic equations (QKE's).
The interaction between different momentum states has been neglected in the
QRE's and it is likely that a full solution of the QKE's would find a somewhat
smaller degree of chaos. However, it also seems well-established that the
asymmetry does indeed oscillate also when the QKE's are solved
\cite{barifoot,wong}, and so one may
expect our analysis to qualitatively also apply to the QKE's.

{\it Acknowledgements ---} One of us (PEB) wishes to thank Nordita, where this work
was completed, for hospitality.

\end{document}